# Combining Named Entities with WordNet and Using Query-Oriented Spreading Activation for Semantic Text Search

Vuong M. Ngo, Tru H. Cao and Tuan M.V. Le

*Faculty of Computer Science and Engineering*
*Ho Chi Minh City University of Technology*
*Viet Nam*
*{vuong.cs@gmail.com}*

**Abstract.** Purely keyword-based text search is not satisfactory because named entities and WordNet words are also important elements to define the content of a document or a query in which they occur. Named entities have ontological features, namely, their aliases, classes, and identifiers. Words in WordNet also have ontological features, namely, their synonyms, hypernyms, hyponyms, and senses. Those features of concepts may be hidden from their textual appearance. Besides, there are related concepts that do not appear in a query, but can bring out the meaning of the query if they are added. We propose an ontology-based generalized Vector Space Model to semantic text search. It exploits ontological features of named entities and WordNet words, and develops a query-oriented spreading activation algorithm to expand queries. In addition, it combines and utilizes advantages of different ontologies for semantic annotation and searching. Experiments on a benchmark dataset show that, in terms of the MAP measure, our model is 42.5% better than the purely keyword-based model, and 32.3% and 15.9% respectively better than the ones using only WordNet or named entities.

## I. INTRODUCTION

There are two types of searches in Information Retrieval (IR) that are Document Retrieval and Question-and-Answering, respectively mentioned as Navigational Search and Research Search in [15]. In practice, answer objects obtained from a Question-and-Answering search engine can be used to search further for documents about them ([12]). Our search engine here is about Document Retrieval, meaning that a user provides the search engine with a phrase or a sentence to look for desired documents.

Much semantic information of documents or queries is lost when each of them is represented by only a set of keywords, as in current search engines like Yahoo or Google. Meanwhile, people often use named entities (NE) in searching. Indeed, in the top 10 search terms by YahooSearch[1] and GoogleSearch[2] in 2008, there are respectively 10 and 9 ones that are NEs. Besides, textual corpora, such as web pages and blogs, often contain NEs.

Named entities are those that are referred to by names such as people, organizations, and locations ([28]) and could be described in ontologies. Each NE may be annotated with its occurring name, class, and identifier if existing in the ontology of discourse. That is, a fully recognized named entity has three features, namely, name, type, and identifier. Due to ambiguity in a context or performance of a recognition method, a named entity may not be fully annotated or may have multiple annotations.

As a popular IR model, the Vector Space Model (VSM) has advantages as being simple, fast, and with a ranking method as good as large variety of alternatives ([2]). However, with general disadvantages of the keyword based IR, the keyword based VSM is not adequate to represent the semantics of queries referring to named entities, for instances: (1) Search for documents about football clubs; (2) Search for documents about Bombay; (3) Search for documents about Paris City; (4) Search for documents about Paris City, Texas, USA.

In fact, the first query searches for documents containing NEs of the class *Football Club*, e.g. *Chelsea, Barcelona* … rather than those containing the keywords "*football club*". For the second query, target documents may mention *Bombay City* under other names, i.e., the city's aliases, such as *Mumbai City*. Besides, documents containing *Bombay Hotel* or *Bombay Company* are also suitable. In the third query, users do not expect to receive answer documents about entities that are also named "*Paris*", e.g. the actress *Paris Hilton*, *University of Paris* but are not cities. Meanwhile, the fourth query requests documents about a precisely identified named entity, i.e., the Paris City in Texas, USA, not the one in France. That is, entity aliases, classes, and identifiers have to be taken into account.

Nevertheless, in many cases, named entities alone do not represent fully the contents of a document or a query. For example, given the query "*earthquake in United States of America*", the keyword "*earthquake*" also conveys important information for searching suitable documents. Besides, there are queries without named entities. Hence, it is necessary to have an IR model that combines NEs and keywords to improve search quality.

Moreover, a word can have other words as its synonyms, hyponyms ... Therefore, with the queries like "*temblor in USA*" or "*natural calamity in USA*", documents about "*earthquake in United States of America*" are truly relevant answers because *earthquake* is a synonym of *temblor* and a hyponym of *natural calamity*. Therfore, a word ontology like WordNet is required for semantic text search.

The other focus of this paper is query expansion with those names entities in an ontology or WordNet words

---

[1] http://buzz.yahoo.com/yearinreview2008/top10/
[2] http://www.google.com/intl/en/press/zeitgeist2008/

(WNs) that are implied by, or related to, the ones in the query. In this paper, we use the term "*concept*" to represent both NEs and WNs. Intuitively, adding correctly related concepts to a query should increase the recall while not sacrificing the precision of searching. For example, given the query to search for documents about "*natural calamity in Southeast Asia*", documents about "*earthquakes in Indonesia or Philippines*" are truly relevant answers, because the two countries are part of *Southeast Asia*.

In this paper, we propose a new ontology-based IR model with three key ideas. First, we propose a NE-WN-based generalized Vector Space Model that uses different ontological features of named entities and WordNet words. Second, the system uses a query-oriented spreading activation algorithm to extract latent concepts relating to the content of a query to expand it. Third, it exploits multiple ontologies to have rich sources for both descriptions and relations of concepts for semantic expansion of documents and queries.

In the next section, we discuss background and related works. Section III describes the proposed system architecture and the methods to expand documents and queries. Section IV presents evaluation of the proposed model and discussion on experiment results in comparison to other models. Finally, section V gives some concluding remarks and suggests future works.

## II. BACKGROUND AND RELATED WORKS

### A. Word Sense Disambiguation Using WordNet

Word sense disambiguation (WSD) is to identify the right meaning of a word in its occurring context. Lesk's algorithm ([20]) was one of the first WSD algorithms for phrases. The main idea of Lesk's algorithm was to disambiguate word senses by finding the overlap among their sense definitions using a traditional dictionary. Then [21] and [3] improved Lesk's algorithm by using WordNet. Following [21], we use associated information of each sense of a word in WordNet, including its definition, synonyms, hyponyms, and hypernyms. By comparing the associated information of each sense of a word with its surrounding words, we can identify its right sense. Each sense has its own identifier.

### B. Spreading Activation

Spreading activation (SA) ([8]) is a method for searching associative networks or semantic networks. The basic idea behind SA is exploitation of relations between nodes in the networks. The nodes may correspond to terms, documents, authors, and so forth. The relations are usually labeled and/or directed and/or weighted. An SA algorithm creates initial nodes that are related to the content of a query and assigns weights to them. After that, the nodes will activate different nodes in a semantic network by some rules and the activated nodes are added to the original query.

### C. Related Works

For Question-and-Answering, [32] presented a method to translate a keyword-based query into a description logic query, exploiting links between named entities in the query. In [6], the target problem was to search for named entities of specified classes associated with keywords in a query, i.e., considering only entity classes for searching. Meanwhile, in [17] the query was converted into SPARQL and the results were ranked by a statistical language model.

For Document Retrieval, the methods in [2] and [24] combined keywords with only NE classes, not considering other features of named entities and combinations of those features. In [5] and [12], a linear combination of keywords and NEs was applied, but a query had to be posted in RDQL to find satisfying NEs before the query vector could be constructed. In [18], it was showed that normalization of entity names improved retrieval quality, which is actually what we call aliases here. As other alternative approaches, [33] and [10] respectively employed WordNet and Wikipedia to expand queries with related terms. In [25], features of named entities were combined with keywords for semantic text search, but WordNet was not used.

Meanwhile, some works expanded queries by using SA algorithms. In [1], the system used a two-level spreading activation network to activate strongly positive and strongly negative matches based on keyword search results. In [26], given an ontology, weights were assigned to links based on certain properties of the ontology, to measure the strength of the links. SA techniques were used to find related concepts in the ontology given an initial set of concepts and corresponding initial activation values. In [13] and [14], the authors mapped the original query to a keyword set and searched for documents relating to the keyword set. The system used an SA algorithm to find concepts relating to those concepts in the related documents, and added them to the original query. In [27], the system found answers of a query to expand it before using an SA algorithm. Generally, those works used all relations of a node in an ontology. Whereas, we use only relations relating to the content of a query. So, our SA algorithm is called a query-oriented one.

## III. ONTOLOGY-BASED INFORMATION RETRIEVAL

### A. System Architecture

Our proposed system architecture of semantic text search is shown in Figure 1. It has two main parts. Part 1 presents document annotation and expansion. Part 2 presents the query expansion module using a query-oriented SA algorithm.

Since no single ontology is rich enough for every domain and application, merging or combining multiple ontologies are reasonable solutions ([7]). Specifically, our proposed model needs an ontology having: (1) a comprehensive class catalog with a large concept population; and (2) many relations between concepts, for expanding queries with latently related entities. So we have combined 3 ontologies, namely, KIM, WordNet, and YAGO to have a rich ontology for both descriptions and relations of concepts for semantic expansion of documents and queries.

In this work we employ KIM ([22]) for automatic NE recognition and semantic annotation of documents and queries. The KIM PROTON ontology contains about 300 classes and 100 attributes and relations. KIM World Knowledge Base (KB) contains about 77,500 entities with more than 110,000 aliases. NE descriptions are stored in an RDF(S) repository. Each NE has information about its specific class, aliases, and attributes (i.e., its own properties or relations with other NEs).

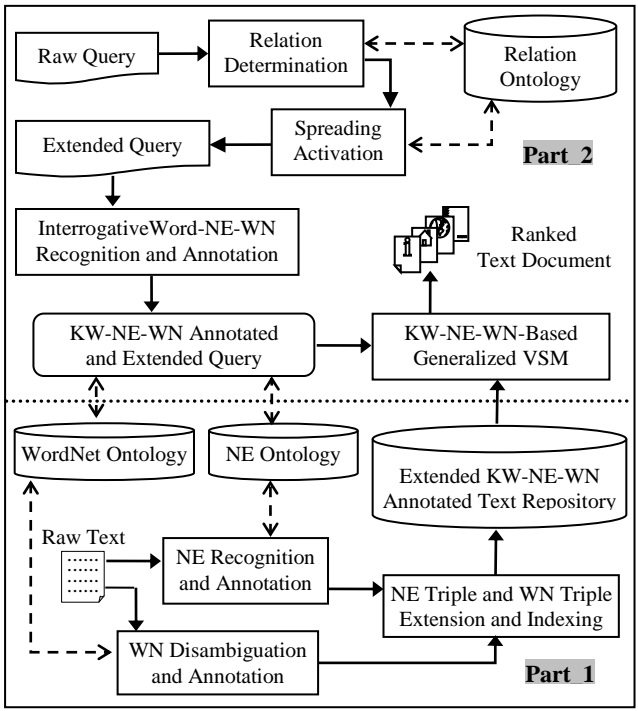

**Fig. 1.** System architecture for semantic text search

WordNet ([35], [11]) is a lexical database for English organized in synonym sets (synsets). There are various semantic relations between these synonym sets, such as hypernym, hyponym, holonym, meronym, and similarity. WordNet contains about 150,000 words organized in over 115,000 synsets. WordNet can be reputed as a lexical ontology. For example, noun synsets have hypernym/ hyponym relationships that can be reputed as relations between concepts in an ontology. We use the algorithm mentioned in section II.A to recognize word senses and embeds semantic annotations in documents and queries.

Since KIM ontology and WordNet define only a small number of relations, and KIM KB contains a limited number of facts, we employ YAGO (Yet Another Great Ontology) ([30], [31]), which is rich in assertions of relations between entities, for an ontology of relations in the system. It contains about 1.95 millions entities, 93 different relation types, and 19 millions facts about specific relations between entities. The facts are extracted from Wikipedia and combined with WordNet using information extraction rules and heuristics. New facts are verified and added to the knowledge base by YAGO core checker. Therefore the correctness of the facts is about 95%. In addition, with logical extraction techniques and a flexible architecture, YAGO can be further extended in future. Note that, to have more relation types and facts, we can employ and combine it with some other ontologies.

The NE Recognition-and-Annotation module and WN Disambiguation-and-Annotation module extract and embed NE triples and WN triples in a raw text, respectively. The text is then indexed by contained NE triples, WN triples, and keywords, and stored in the Extended KW-NE-WN Annotated Text Repository. Meanwhile, the InterrogativeWord-NE-WN Recognition-and-Annotation module extracts and embeds the most specific NE triples, WN triples in the extended query, and replaces the interrogative word if existing by a suitable class. Semantic document search is performed via the KW-NE-WN-Based Generalized VSM module.

### B. Query and Document Concept-Based Annotation and Expansion

We propose a generalized VSM in which a document or a query is represented by a vector over a space of generalized terms. Each term is a NE triple, a WN triple, or a keyword. As usual, similarity of a document and a query is defined by the cosine of the angle between their representing vectors. Our work has implemented the model by developing a platform called S-Lucene modified from Lucene[3]. The system automatically processes documents for KW-NE-WN-based searching in the following steps:

1. Removing stop-words in the documents.
2. Recognizing and annotating NEs in the documents using KIM[4].
3. Disambiguating and annotating WordNet words that are not NEs in the document using WordNet and the WSD algorithm mentioned in section II.A.
4. Extending the documents with implied NE triples. That is, for each entity named *n* possibly with class *c* and identifier *id* in a document, the triples (*n/\*/\**), (*\*/c/\**), (*n/c/\**), (*alias*(*n*)/*\*/\**), (*\*/super*(*c*)/*\**), (*n/super*(*c*)/*\**), (*alias*(*n*)/*c/\**), (*alias*(*n*)/ *super*(*c*)/*\**), and (*\*/\*/id*) are virtually added to the document.
5. Extending the documents with implied WN triples. That is, for each WordNet word having a sense named *w* with hypernym *h* and identifier *id* in a document, the triples (*w/\*/\**), (*\*/h/\**), (*w/h/\**), (*syn*(*w*)/*\*/\**), (*\*/super*(*h*)/*\**), (*w/super*(*h*)/*\**), (*syn*(*w*)/*h/\**), (*syn*(*w*)/*super*(*h*)/*\**), and (*\*/\*/id*) are virtually added to the document.
6. Indexing NE triples, WN triples, and keywords by S-Lucene.

Here *alias*(*n*), *super*(*c*), *syn*(*w*) and *super*(*h*) respectively denote any alias of *n*, any super class of *c*, any synonym of $w$, and any super hypernym of $h$ in the ontology and knowledge base of discourse.

---

[3] http://lucene.apache.org/

[4] http://www.ontotext.com/kim/

A query is also automatically processed in the following steps:
1. Removing stop-words in the query.
2. Recognizing and annotating named entities in the query.
3. Recognizing and annotating WordNet words that are not NEs in the query.
4. Representing each recognized entity named $n$ possibly with class $c$ and identifier $id$ by the most specific and available triple among ($n$/*/*), (*/$c$/*), ($n$/$c$/*), and (*/*/$id$).
5. Representing each recognized WordNet word having a sense named $w$ with hypernym $h$ and identifier $id$ by the most specific and available triple among ($w$/*/*), (*/$h$/*), ($w$/$h$/*), and (*/*/$id$).

Besides, there is latent information of the interrogative words *Who*, *What*, *Which*, *When*, *Where*, or *How* in a query. For example, given the query "*Where was George Washington born?*", the important terms are not only the NE *George Washington* and the word "*born*", but also the interrogative word *Where*, which is to search for locations or documents mentioning them. For instance, *Where* in this example should be mapped to the class *Location*. The mapping could be automatically done with high accuracy using the method proposed in [4].

### C. Query Expansion Using a Query-Oriented SA Algorithm

The following are the four main steps of our method to determine latently related concepts for a query:
1. Recognizing Relation Phrases: Relation phrases are prepositions, verbs, and other phrases representing relations, such as *in*, *on*, *of*, *has*, *is*, *are*, *live in*, *located in*, *was actress in*, *is author of*, *was born*. We have implemented a relation phrase recognition using the ANNIE tool of GATE ([9]).
2. Determining Relations: Each relation phrase recognized in step 1 is mapped to the corresponding one in an ontology of relations by a manually built dictionary. For example, "*was actress in*" is mapped to *actedIn*, "*is author of*" is mapped to *wrote*, and "*nationality is*" is mapped to *isCitizenOf*.
3. Recognizing Concepts: Recognizing NEs and WordNet words with the same tools used in the document annotation process.
4. Determining Related Concepts: Each concept that is combined with a relation determined in step 2 and a concept recognized in step 3 is added to the query. The weight of the activated concept is equal to the weight of the original concept. In the scope of this paper, we consider to expand only queries having one relation. However, the method can be applied straightforwardly to queries with more than one relation.

## IV. EXPERIMENTS

### A. Dataset and Performance Measures

Evaluation of a retrieval model or method requires two components: (1) a test dataset including one document collection, one query collection and relevance information about which documents are relevant to each query; and (2) quality measures based on relevant and non-relevant documents returned for each query ([2], [23]).

We choose the L.A. Times document collection of TREC datasets, since there are 59% full papers about text IR of SIGIR-2007[5] and SIGIR-2008[6] using that document collection. The L.A. Times consists of more than 130,000 documents in nearly 500MB. Next, we choose queries of the QA-Track that have answer documents in this document collection. As presented in Table 1, among the queries of the QA-Track, there are only 756 queries of QA-99, QA-00, QA-01 that have answer in the L.A. Times. Within these 756 queries, there are 51 queries each of which contains a single relation phrase that has a respective relation type and facts in the ontology of discourse. These are the queries that could be expanded for testing experiments.

**Table 1.** Query survey of the TREC QA-Track

| QA-Track | | 1999 | 2000 | 2001 | Total |
|---|---|---|---|---|---|
| Number of queries | | 200 | 693 | 500 | 1393 |
| Number of queries having answers in LA-Times | | 124 | 403 | 229 | 756 |
| Number of queries having answers in LA-Times | Queries with more than one relation phrase | 55 | 38 | 3 | 96 |
| | Queries with only one relation phrase (ORP) | 69 | 365 | 226 | 660 |
| | ORP + the relation phrase has a relation type (ORPT) | 33 | 112 | 53 | 198 |
| | **ORPT + there are respective facts in the ontology** | **17** | **11** | **23** | **51** |

We have evaluated and compared the new models in terms of precision-recall (P-R) curves, F-measure-recall (F-R) curves, and single mean average precision (MAP) values. The average P-R curve and average F-R curve over all the queries are obtained by P-R curves and F-R curves of each query which are interpolated to the eleven standard recall levels that are 0%, 10%, …, 100%, as in [19] and [23]. Meanwhile, MAP is a single measure of retrieval quality across recall levels and considered as a standard measure in the TREC community ([34]).

### B. Statistical Significance Testing

Obtained values of the measures presented above might occur by chance due to: (1) the specific and restricted contents of queries and documents in a test dataset; (2) the subjective judgment of human assessors on relevance between the test queries and documents; and (3) the limited numbers of queries and documents used in an evaluation experiment. So, when comparing systems, a typical null hypothesis is that they are equivalent in terms of performance, even though their quality measures appear different. In order to reject this null hypothesis and confirm that one system truly performs better than another, a statistical significance test is required ([16]).

---

[5] http://www.sigir2007.org
[6] http://www.sigir2008.org

In [29], the authors compared the five significant tests that have been used by researchers in information retrieval, namely, Student's paired t-test, Wilcoxon signed rank test, sign test, bootstrap, and Fisher's randomization (permutation). They recommended Fisher's randomization test for evaluating the significance of the observed difference between two systems. In practice, it is usually infeasible to compute an exponential number of permutations. As shown in [29], 100,000 permutations were acceptable for a randomization test and the threshold 0.05 of the two-sided significance level, or p-value, could detect significance.

**Table 2.** The average precisions and F-measures at the eleven standard recall levels on 51 expandable queries of TREC

| Measure | Model | Recall (%) | | | | | | | | | | |
|---|---|---|---|---|---|---|---|---|---|---|---|---|
| | | 0 | 10 | 20 | 30 | 40 | 50 | 60 | 70 | 80 | 90 | 100 |
| Precision (%) | Keyword | 57 | 57 | 55 | 50 | 45 | 45 | 34 | 28 | 26 | 24 | 24 |
| | WordNet | 59 | 59 | 57 | 53 | 48 | 48 | 37 | 31 | 30 | 28 | 28 |
| | KW+NE+Wh | 63 | 63 | 61 | 57 | 53 | 52 | 45 | 40 | 38 | 36 | 36 |
| | Semantic Search | 69 | 69 | 67 | 63 | 60 | 59 | 53 | 48 | 47 | 46 | 45 |
| F-measure (%) | Keyword | 0 | 15 | 25 | 32 | 35 | 39 | 35 | 33 | 32 | 31 | 32 |
| | WordNet | 0 | 15 | 25 | 32 | 36 | 41 | 37 | 34 | 35 | 34 | 34 |
| | KW+NE+Wh | 0 | 15 | 26 | 34 | 39 | 44 | 43 | 43 | 43 | 42 | 43 |
| | Semantic Search | 0 | 16 | 28 | 37 | 43 | 48 | 49 | 47 | 49 | 51 | 52 |

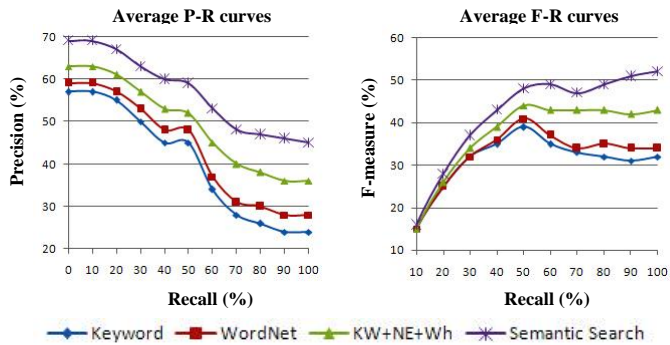

**Fig. 2.** Average P-R and F-R curves of Keyword, WordNet, KW+NE+Wh and Semantic search models on 51 queries of TREC

**Table 3.** The mean average precisions on the 51 queries of TREC

| Model | Keyword | WordNet | KW+NE+Wh | Semantic Search |
|---|---|---|---|---|
| MAP | 0.3934 | 0.4236 | 0.4836 | 0.5604 |

**Table 4.** Randomization tests of Semantic Search with the Keyword, WordNet, and KW+NE+Wh models

| Model A | Model B | \|MAP(A) − MAP(B)\| | N⁻ | N⁺ | Two-Sided P-Value |
|---|---|---|---|---|---|
| Semantic Search | Keyword | 0.1670 | 1531 | 1572 | 0.0310 |
| | WordNet | 0.1369 | 2337 | 2499 | 0.0484 |
| | KW+NE+Wh | 0.0768 | 5299 | 5472 | 0.1077 |

*C. Testing Results*

We conduct experiments to compare the results obtained by four different search models:
1. Keyword Search: This search uses Lucene text search engine as a tweak of the traditional keyword-based VSM.
2. WordNet Search: This search uses only WordNet for word sense disambiguation and annotation of documents and queries.
3. KW+NE+Wh Search: This search uses only a NE ontology for NE recognition and annotation of documents and queries. In addition, interrogative words of queries are replaced by suitable classes.
4. Semantic Search: This search uses the proposed model and system presented in section III.

Table 2 presents, and Figure 2 plots, the average precisions and F-measures of the Keyword, WordNet, KW+NE+Wh, and Semantic search models at each of the standard recall levels. It shows that Semantic Search performs better than the other three models, in terms of the precision and F measures. Figure 3 shows the per query differences in average precision of Semantic Search with the Keyword, WordNet, and KW+NE+Wh models. The MAP values in Table 3 and the two-sided p-values in Table 4 show that taking into account latent ontological features and related concepts in queries and documents does enhance text retrieval performance. In terms of the MAP measure, Semantic Search performs about 42.5% better than the Keyword model, and about 32.3% and 15.9% better than the WordNet and KW+NE+Wh models, respectively.

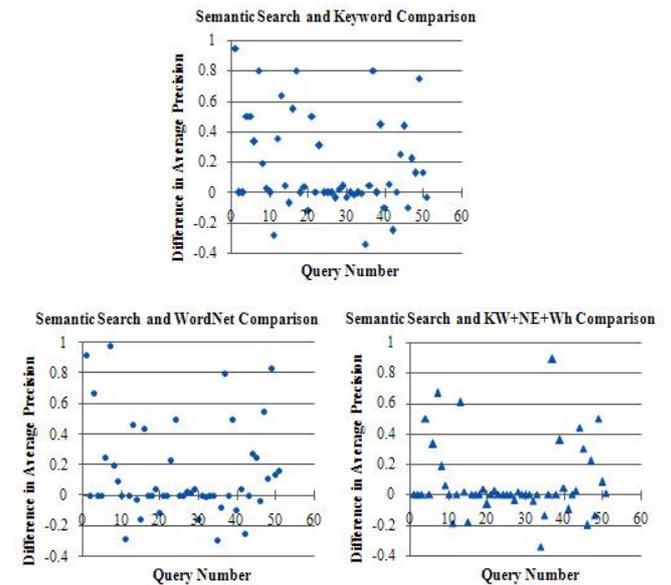

**Fig. 3.** The per query differences in average precision of Semantic Search with Keyword, WordNet, KW+NE+Wh models

## V. CONCLUSION AND FUTURE WORKS

We have presented the generalized VSM that exploits ontological features of named entities and WordNet words for semantic text search. That is a whole IR process, from a natural language query to a set of ranked answer documents. Besides, given an ontology, we use a query-oriented SA algorithm to exploit latent concepts relating to original concepts in a query and enrich the query with them. We have also proposed a framework to combine multiple ontologies to take their full advantages for the whole semantic search process.

The conducted experiments on a TREC dataset have showed that our appropriate NE ontology and WordNet exploitation improves the search quality in terms of the precision, recall, F, and MAP measures. For future work, we are considering combination with more ontologies to increase the relation coverage and researching methods to better recognize relations in a query.